\documentclass{elsart}

\usepackage[english]{babel}

\usepackage{graphicx}
\usepackage{amsmath,amssymb}
\usepackage[square,comma,sort&compress]{natbib}

\newcommand{\pnu}[1] {\overset{\smash{\scriptscriptstyle (-)}}{\nu}_{\hskip-3pt #1}}

\DeclareMathAlphabet{\mathsc}{OT1}{cmr}{m}{sc}

\newcommand{\CL} {C.L.}
\newcommand{\dof}{d.o.f.}
\newcommand{\gof}{g.o.f.}
\newcommand{\eVq}{\text{eV}^2}

\newcommand{\Sol}  {\mathsc{sol}}
\newcommand{\Atm}  {\mathsc{atm}}
\newcommand{\Sbl}  {\mathsc{sbl}}
\newcommand{\Nev}  {\mathsc{nev}}
\newcommand{\Lsnd} {\mathsc{lsnd}}

\newcommand{\dms}{\Delta m^2_\Sol}
\newcommand{\dma}{\Delta m^2_\Atm}
\newcommand{\dml}{\Delta m^2_\Lsnd}
\newcommand{\thl}{\theta_\Lsnd}

\begin{document}
\raisebox{8mm}[0pt][0pt]{\hspace{12cm}
\vbox{hep-ph/0207157\\IFIC/02-28\\UWThPh-2002-20}}

\begin{frontmatter}
 
\title{\vspace*{1 cm} Ruling out four-neutrino oscillation  %
    interpretations of the LSND anomaly?}                   %
\author{M.~Maltoni}$^a$, \ead{maltoni@ific.uv.es}           %
\author{T.~Schwetz}$^b$, \ead{schwetz@thp.univie.ac.at}     %
\author{M.~A.~T\'ortola}$^a$ and \ead{mariam@ific.uv.es}    %
\author{J.~W.~F.~Valle}$^a$ \ead{valle@ific.uv.es}          %
\address{$^a$Instituto de F\'{\i}sica Corpuscular --        %
  C.S.I.C./Universitat de Val{\`e}ncia \\                   %
  Edificio Institutos de Paterna, Apt 22085,                %
  E--46071 Valencia, Spain\\
  $^b$Institut f\"ur Theoretische Physik, Universit\"at Wien \\
  Boltzmanngasse 5, A-1090 Wien, Austria\vspace{2 cm}}      %

\begin{abstract}
  Prompted by recent solar and atmospheric data, we re-analyze the
  four-neutrino description of current global neutrino oscillation data,
  including the LSND evidence for oscillations. 
  The higher degree of rejection for non-active solar and atmospheric
  oscillation solutions implied by the SNO neutral current result as
  well as by the latest 1489-day Super-K atmospheric neutrino data
  allows us to rule out (2+2) oscillation schemes proposed to
  reconcile LSND with the rest of current neutrino oscillation data.
  Using an improved goodness of fit (\gof) method especially sensitive
  to the combination of data sets we obtain a \gof\ of only $1.6\times
  10^{-6}$ for (2+2) schemes. Further, we re-evaluate the status of
  (3+1) oscillations using two different analyses of the LSND data
  sample. We find that also (3+1) schemes are strongly disfavoured by
  the data. Depending on the LSND analysis we obtain a \gof\ of
  $5.6\times 10^{-3}$ or $7.6\times 10^{-5}$. This leads to the
  conclusion that all four-neutrino descriptions of the LSND anomaly,
  both in (2+2) as well as (3+1) realizations, are highly disfavoured.
  Our analysis brings the LSND hint to a more puzzling status.
\begin{keyword}
neutrino oscillations \sep sterile neutrino \sep four-neutrino models
\PACS 14.60.P \sep 14.60.S \sep 96.40.T \sep 26.65 \sep 96.60.J \sep 24.60
 \end{keyword}
\end{abstract}
\end{frontmatter}

\section{Introduction}

The atmospheric neutrino data~\cite{atm-exp,skatm,macro}, including
the most recent 1489 Super-K data sample provide strong evidence for
$\nu_\mu$ oscillations into an active neutrino (mainly $\nu_\tau$)
with $\dma\sim 2 \times 10^{-3}~\eVq$~\cite{solat02}.
On the other hand, apart from confirming, once again, the
long-standing solar neutrino
problem~\cite{sksol,chlorine,sage,gallex_gno,sno01}, the recent
results from the Sudbury Neutrino Observatory (SNO) \cite{sno02} have
given strong evidence that solar neutrinos convert mainly to an active
neutrino flavor.  This suggests that an extension of the Standard
Model of particle physics is necessary in the lepton sector, capable
of incorporating the required $\nu_e$ conversion.
Although certainly not yet unique \cite{Miranda:2000bi,Guzzo:2001mi},
the most popular explanation of the solar neutrino data is provided by
the active large mixing angle (LMA) neutrino oscillation hypothesis,
characterized by a neutrino mass-squared difference $\dms\lesssim
10^{-4}~\eVq$~\cite{solat02}.

In contrast, reactor and accelerator neutrino data
\cite{KARMEN,bugey,CHOOZ,PaloV,CDHS} lead to negative results.
However, the LSND experiment~\cite{LSND,LSND2001} has provided
positive results, which may or may not be confirmed by the
forth-coming MiniBooNE experiment~\cite{MiniBooNE}.  The required
neutrino mass-squared difference $\dml \gtrsim 0.2$ eV$^2$ is in
conflict with the neutrino mass-squared differences indicated by solar
and atmospheric data in a minimal three-neutrino picture.
Four-neutrino models
\cite{ptv92,pv93,cm93,4nuModels,4nuextra,Ioannisian:2001mu,Hirsch:2000xe}
potentially account for all current oscillation data.
The status of four-neutrino descriptions has been presented in
Ref.~\cite{Maltoni:2001bc}. An exhaustive list of four-neutrino references can
be found in Ref.~\cite{giuntiwebp}.

\begin{figure}[t]
 \centering
   \includegraphics[width=0.65\linewidth]{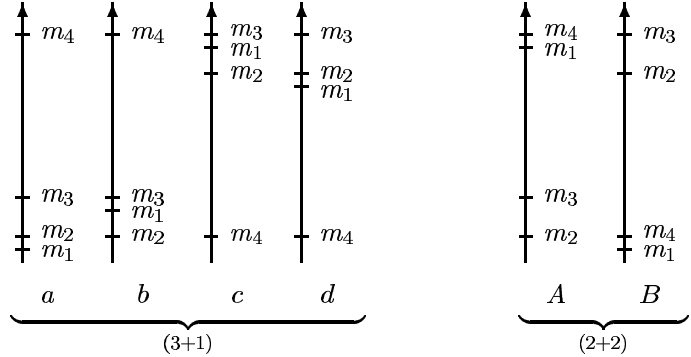}
      \caption{\label{fig:4spectra}%
        The six types of four-neutrino mass spectra. The spacings in
        the vertical axes correspond to the different scales of
        mass-squared differences required in solar, atmospheric and
        short baseline neutrino oscillations.}
\end{figure}

The six possible four-neutrino mass spectra are shown in
Fig.~\ref{fig:4spectra}.  For the case $\dml \gg \dma, \dms$, which we
tacitly assume in order to reconcile the LSND anomaly with solar and
atmospheric data, these schemes can naturally be divided into two
very different classes, usually called (3+1) and (2+2) mass
schemes~\cite{barger00}.
It is important to note that (3+1) mass spectra include the
three-active neutrino scenario as limiting case. In this case solar
and atmospheric neutrino oscillations are explained by active neutrino
oscillations, with mass-squared differences $\dms$ and $\dma$, and the
fourth neutrino state gets completely decoupled.  We will refer to
such limiting scenario as (3+0).  In contrast, the (2+2) spectrum is
intrinsically different, as there must be a significant contribution
of the sterile neutrino either in solar or in atmospheric neutrino
oscillations or in both.

In recent studies~\cite{Maltoni:2001bc} it has been realized that
there is considerably tension in four-neutrino fits to the global
data. In such global four-neutrino analyses one is faced with the
problem that there are different data sets, which all give very good
fits if analyzed separately. Problems arise due to the {\sl
  combination} of the different data sets in a four-neutrino
framework.  In this letter we re-evaluate the status of four-neutrino
interpretations of the LSND anomaly in the light of the recent solar
\cite{sksol,sage,sno02} as well as atmospheric \cite{skatm} neutrino
results.  To evaluate the quality of the fit we will apply appropriate
statistical methods, which are especially sensitive to the combination
of different data sets in a global analysis.
We find that both the SNO NC result as well as the 1489-day Super-K
atmospheric neutrino data strongly reject against sterile neutrino
conversions. This essentially rules out (2+2) mass schemes. We also
re-evaluate the status of (3+1) schemes by considering two different
analyses of LSND data~\cite{LSND2001,Church:2002tc}.  
We will further elaborate the result of previous studies
\cite{barger00,3+1early,BGGS,peres,carlo,GS,Maltoni:2001mt} that in
(3+1) schemes the LSND signal is in serious disagreement with bounds
from short-baseline (SBL) experiments reporting no evidence for
oscillations \cite{KARMEN,bugey,CDHS}.
The net result is that all four-neutrino descriptions of the LSND
anomaly, both in (2+2) as well as (3+1) realizations, are strongly
disfavoured by the data. This brings the LSND anomaly to a
theoretically more puzzling status. We note that also cosmology put
strong constraints on four-neutrino schemes (for recent analyses see
Ref.~\cite{cosmology}).

The plan of the paper is as follows. In
Sec.~\ref{sec:four-neutr-oscill} we briefly describe our
parameterization and approximations used in the global four-neutrino
analysis.  In Secs.~\ref{sec:solar} and \ref{sec:atmosph} we summarize
the solar and atmospheric neutrino data and their analysis
\cite{solat02}. In Sec.~\ref{sec:sbl} we describe the SBL data we are
using, and we compare the two different analyses of the LSND
data~\cite{LSND2001,Church:2002tc}.
In Sec.~\ref{sec:2+2} we show that (2+2) oscillation schemes are ruled
out because of the tension between solar and atmospheric data, whereas
in Sec.~\ref{sec:3+1} we present our re-analysis of the disagreement
in SBL data in (3+1) oscillation schemes in light of the two LSND
samples.  In Sec.~\ref{sec:comparing} we show that the goodness of fit
is very bad in all four-neutrino cases. Furthermore, we compare the
relative quality of the fit for the schemes (3+1) and (2+2), as well
as the three active neutrino case (3+0).  The quantitative statistical
criteria to analyze the conflict between different data sets, which
are used in Secs.~\ref{sec:2+2}, \ref{sec:3+1} and
\ref{sec:comparing}, are formulated in the appendix \ref{appendix}.
In summary, we find that all four-neutrino descriptions of the LSND
anomaly both in (2+2) as well as (3+1) realizations are highly
disfavoured due to recent data, as mentioned in our conclusions,
Sec.~\ref{sec:conclusions}.

\section{Four-neutrino oscillations}
\label{sec:four-neutr-oscill}

In this letter we interpret current neutrino oscillation data in a
four-neutrino framework. This requires the existence of a fourth light
neutrino, which must be sterile to account for the LEP invisible Z
width data~\cite{ptv92,pv93,cm93,4nuModels,giuntiwebp}. The simplest
theoretical models add just one such light $SU(2) \otimes U(1)$
singlet neutrino and ascribe its lightness either to an underlying
protecting symmetry, such as lepton number~\cite{ptv92,pv93}, or to
the existence of extra dimensions~\cite{4nuextra,Ioannisian:2001mu}.
Symmetry breaking is required in order to generate solar and
atmospheric oscillations, characterized by splittings which arise due
to small tree--level \cite{Hirsch:2000xe} or radiative effects
\cite{ptv92,pv93}.  Many other four-neutrino models have been
considered \cite{4nuModels,giuntiwebp}.

Quite generally, the four-neutrino charged current leptonic weak
interaction is specified as by a rectangular $3\times 4$ lepton mixing
matrix $K = \Omega U$, with $\Omega$ diagonalizing the $3\times 3$
charged lepton mass matrix and $U$ diagonalizing the $4\times 4$
Majorana neutrino mass matrix. 
Altogether the elements of the mixing matrix $K$ are characterized by
6 mixing angles and 3 physical phases \cite{Schechter:1980gr}, which
could lead to CP violation in oscillation
phenomena~\cite{Schechter:1981gk}.  For simplicity we neglect
CP-violating phases, whose effects are expected to be small in the
experiments we consider, due to the strong hierarchy of the
mass-squared differences required by the experimental data.
This leaves us with nine parameters altogether relevant for the
description of CP conserving four-neutrino oscillations: 6 mixing
angles contained in the mixing matrix and 3 mass-squared differences.

As shown in Ref.~\cite{Maltoni:2001bc} it is very important to adopt a
convenient choice for these parameters in order to factorize the
analysis. Here we adopt all the notation and conventions introduced in
Ref.~\cite{Maltoni:2001bc}, which are based on the use of physically
relevant quantities.  The 6 parameters $\dms$, $\theta_\Sol$, $\dma$,
$\theta_\Atm$, $\dml$, $\theta_\Lsnd$ are similar to the 2-neutrino
mass-squared differences and mixing angles and are directly related to
the oscillations in solar, atmospheric and the LSND experiments. For
the remaining 3 parameters we use $\eta_s,\eta_e$ and $d_\mu$. Here,
$\eta_s \,(\eta_e)$ is the fraction of $\nu_s \,(\nu_e)$ participating
in solar oscillations, and $1-d_\mu$ is the fraction of $\nu_\mu$
participating in atmospheric oscillations (for exact definitions see
Ref.~\cite{Maltoni:2001bc}).  For the analysis we adopt the following
approximations:
\begin{itemize}
\item 
We make use of the hierarchy 
$\dms \ll \dma \ll \dml$.
This means that for each data set we consider only one mass-squared
difference, the other two are set either to zero or to infinity.
\item
In the analyses of solar and atmospheric data (but not for SBL data) we
set $\eta_e = 1$, which is justified because of strong constraints
from reactor experiments~\cite{bugey,CHOOZ,PaloV}.
\end{itemize}

\begin{figure}[t]
 \centering
   \includegraphics[width=0.65\linewidth]{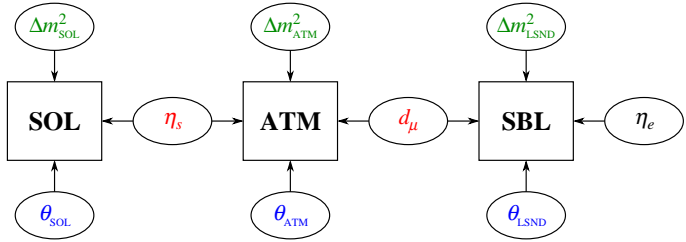}
      \caption{\label{fig:diagram}%
         Parameter dependence of the different data sets in our
         parameterization. }
\end{figure}

Due to these approximations (well justified by reactor data and in
oder to fit in LSND with solar + atmospheric data) the parameter
structure of the problem gets rather simple \cite{Maltoni:2001bc}. The
parameter dependence of the different data sets solar, atmospheric and
SBL is illustrated in Fig.~\ref{fig:diagram}. We see that only
$\eta_s$ links solar and atmospheric data and $d_\mu$ links
atmospheric and SBL data.  All the other parameters are ``private'' to
one data set.

\section{ Solar neutrinos}
\label{sec:solar}

Here we borrow from Ref.~\cite{solat02} a recent re-analysis of
present solar data including a study of degree of rejection against
the sterile neutrino oscillation hypothesis. The experimental data
used are the solar neutrino rate of the chlorine experiment
Homestake~\cite{chlorine}, the rates of the gallium experiments
SAGE~\cite{sage}, GALLEX and GNO~\cite{gallex_gno}, as well as the
1496-day Super-Kamiokande data sample~\cite{sksol} in form of 44
recoil electron zenith-energy bins.
Moreover, we include the data from the recent SNO charged current
and neutral current event measurements, as well as day-night information
in the form of 34 data points of the observed energy spectrum \cite{sno02}.
For the solar neutrino fluxes we use the Standard Solar Model fluxes
\cite{ssm}, including its prediction of the $^8$B flux.
The first hints in favour of the LMA solution
\cite{Gonzalez-Garcia:1999aj}, which followed mainly from the flatness
of the Super-K spectrum, have now become a robust result, thanks to
the new SNO data and additional Super-K spectrum data. In contrast
with the pre-SNO-NC situation, presently all non-LMA solutions are
strongly rejected, so that in this paper we will restrict the solar
oscillation parameters to the LMA region.

\begin{figure}
  \centering 
  \includegraphics[width=0.65\linewidth]{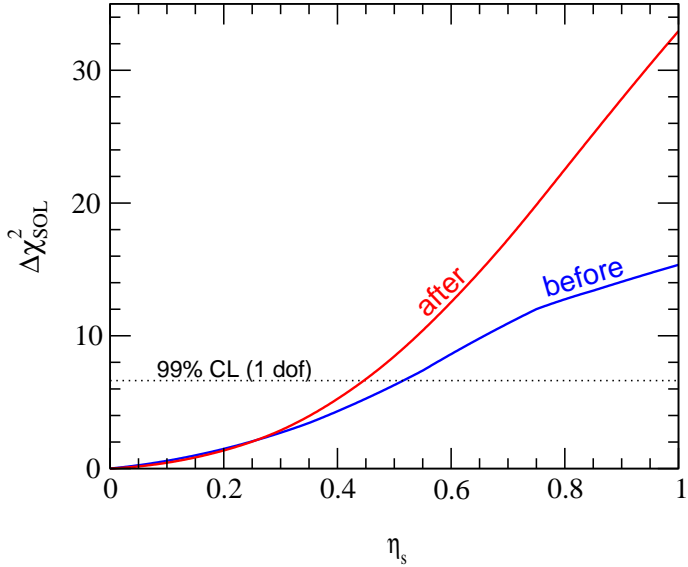}
  \caption{$\Delta \chi^2_\Sol$ for the favoured LMA solution,
    before and after SNO NC data \cite{sno02},
    displayed as a function of $\eta_s$.}
  \label{fig:sol02.etas}
\end{figure}

Another important consequence of the recent SNO-NC data is specially
relevant for us, namely, that the degree of rejection against solar
neutrino oscillations to sterile neutrinos has now become much more
significant. This results can be summarized in Fig.
\ref{fig:sol02.etas}. The line labeled ``after'' shows
$\Delta\chi^2_\Sol$ as a function of $0 \leq \eta_s \leq 1$, after
recent solar neutrino up-dates, most important SNO-NC~\cite{sno02}. We
minimize with respect to the other two solar neutrino oscillation
parameters $\dms$ and $\theta_\Sol$, required only to lie in the LMA
region. For comparison we show also the line labeled ``before'',
corresponding to the data used in Ref.~\cite{Maltoni:2001bc}. From
this figure one can place a mass scheme-independent limit on the
parameter $\eta_s$ from solar data:
\begin{equation}\label{eq:etasSol}
    \text{solar data (LMA):}\quad \eta_s \leq 0.45 \qquad\mbox{(99\% \CL)},
\end{equation}
compared to $\eta_s \leq 0.52$ obtained in Ref.~\cite{Maltoni:2001bc}. Moreover, 
the pure sterile solution ($\eta_s=1$) and large values of $\eta_s$
are now much more disfavoured than previously.

\section{Atmospheric neutrinos}
\label{sec:atmosph}

For the atmospheric data analysis~\cite{solat02} we use the following
data from the Super-Kamiokande experiment~\cite{skatm}: $e$-like and
$\mu$-like samples of sub- and multi-GeV data, as well as the up-going
stopping and through-going muon data. The Super-K data correspond to
the 1489-day sample presented at Neutrino 2002 \cite{skatm}.
In contrast to previous analyses \cite{Maltoni:2001bc,GMPV,Concha2+2}
here we adopt a more precise ten-bin presentation of the corresponding
contained zenith-angle distributions.  Further, we use the data from
MACRO~\cite{macro}, including the recent update of their up-going muon
sample (10 angular bins). Since $\dms \ll \dma$ we adopt the standard
approximation of neglecting $\dms$ when performing the analysis of the
atmospheric data\footnote{For detailed studies of this issue see
  Ref.~\cite{Gonzalez-Garcia:2002mu}.}.  For further details of the
atmospheric neutrino analysis see Refs.~\cite{GMPV,Concha2+2} and
references therein.

Let us consider the impact of atmospheric data alone on the parameter
$\eta_s$.  As shown in Ref.~\cite{Maltoni:2001bc} the behaviour of the
atmospheric data as a function of $\eta_s$ depends on the mass schemes
considered, with three inequivalent cases\footnote{Note that $\eta_s$
is defined as the fraction of sterile neutrinos in {\sl solar}
oscillations and is only indirectly related to the fraction of sterile
neutrinos in {\sl atmospheric} oscillations. This relation is
different for the various mass schemes \cite{Maltoni:2001bc}.}
(3+1)$_{a,d}$, (3+1)$_{b,c}$ and (2+2)$_{A,B}$. The qualitative
behaviour of (3+1)$_{a,d}$ and (3+1)$_{b,c}$ schemes (see
Fig.~\ref{fig:4spectra}) is similar, however, small differences appear
due to matter effects in atmospheric oscillations. In the following we
will always minimize between these two cases and refer simply to the
(3+1) scheme.

\begin{figure}[t]
  \centering 
  \includegraphics[width=0.85\linewidth]{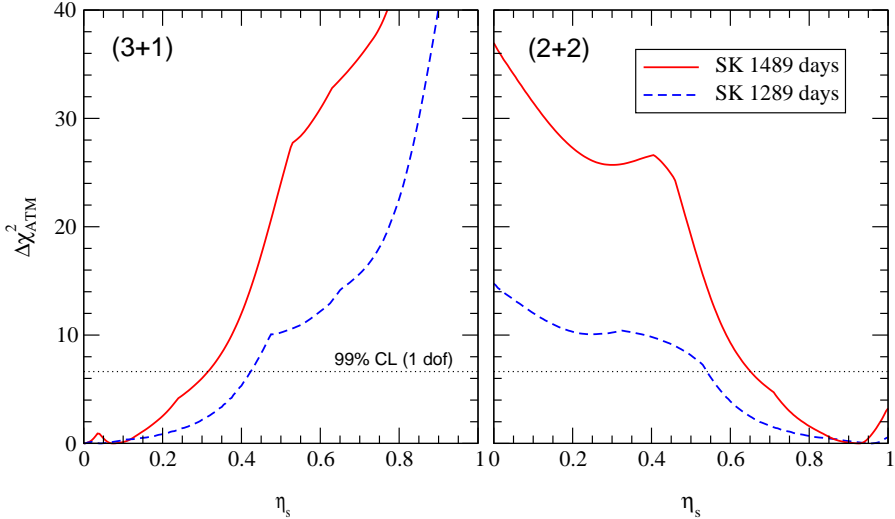}
  \caption{$\Delta\chi^2_\Atm$ displayed as a
    function of $\eta_s$ for (3+1) and (2+2) schemes, before and after
    the most recent Super-K up-date \cite{skatm}.}
  \label{fig:atm02.etas}
\end{figure}

We present in Fig.~\ref{fig:atm02.etas} the $\Delta\chi^2_\Atm$
profile displayed as a function of $0 \leq \eta_s \leq 1$ minimized
with respect to all other parameters, before and after the latest data
of Ref.~\cite{skatm} have been included.  We see that the new data
significantly improve the sensitivity of atmospheric data with respect
to a sterile component. Especially for (2+2) schemes small values of
$\eta_s$ are highly disfavoured, which will be crucial for the
exclusion of (2+2) spectra.  From the figure we obtain the following
bounds on $\eta_s$ at 99\% \CL:
\begin{equation} \label{eq:etasAtm}
    \text{atmospheric data:}
    \begin{cases}
        \eta_s \leq 0.32 & \text{for (3+1) schemes}, \\
        \eta_s \geq 0.65 & \text{for (2+2) schemes}.
    \end{cases}
\end{equation}

\section{Reactor and accelerator data}
\label{sec:sbl}

We divide the data from reactor and accelerator experiments in data
from experiments reporting no evidence (NEV) for oscillations and in
the LSND experiment. In this paper we compare the implications of two
different analyses of the LSND data. First, we use the likelihood
function obtained in the final LSND analysis \cite{LSND2001} from
their global data with an energy range of $20 < E_e < 200$ MeV and no
constraint on the likelihood ratio $R_\gamma$ (see
Ref.~\cite{LSND2001} for details). This sample contains 5697 events
including decay-at-rest (DAR) $\bar\nu_\mu\to\bar\nu_e$, and
decay-in-flight (DIF) $\nu_\mu\to\nu_e$ data. We refer to this
analysis as {\sl LSND global}. Second, we use the LSND analysis
performed in Ref.~\cite{Church:2002tc} based on 1032 events obtained
from the energy range $20 < E_e < 60$ MeV and applying a cut of
$R_\gamma > 10^{-5}$. These cuts eliminate most of the DIF events from
the sample, leaving mainly the DAR data, which are more sensitive to
the oscillation signal. We refer to this analysis as {\sl LSND DAR}.

\begin{figure}[t]
  \centering 
  \includegraphics[width=0.75\linewidth]{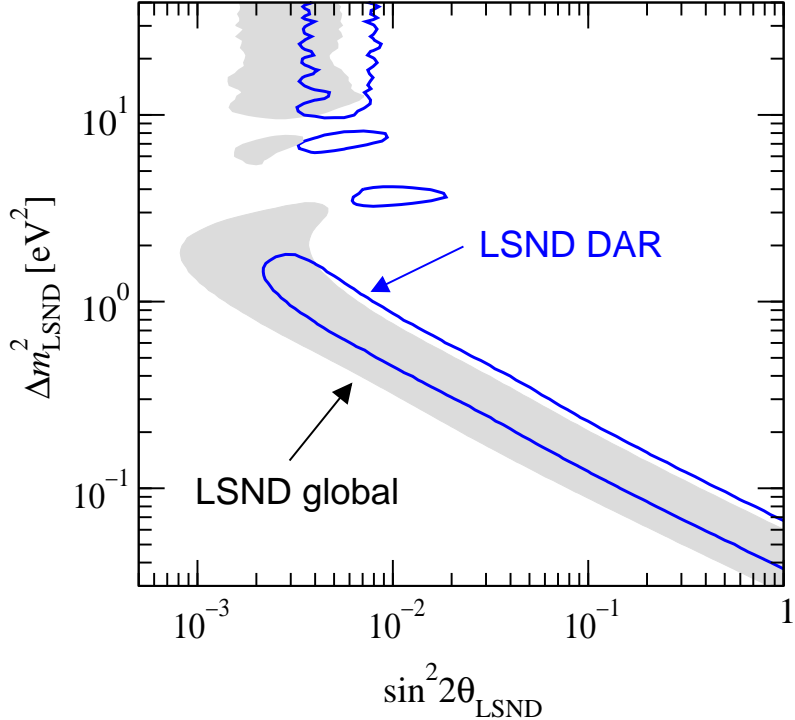}
  \caption{Comparison of the 99\% \CL\ regions of the global LSND
  analysis \cite{LSND2001} (shaded region) and of the analysis performed
  in Ref.~\cite{Church:2002tc} (solid line) using mainly the
  decay-at-rest (DAR) data sample.}
  \label{fig:lsnd}
\end{figure}

In both cases the collaboration has provided us with the likelihood
function obtained in their analyses \cite{LSNDpriv,eitel} which we
convert into a $\chi^2$ according to $\chi^2\propto -2\ln \mathcal{L}$
(see Ref.~\cite{Maltoni:2001bc} for details).  In Fig.~\ref{fig:lsnd}
we compare the 99\% \CL\ regions obtained from the two LSND
analyses. The LSND DAR data prefers somewhat
larger mixing angles, which will lead to a stronger disagreement of
the data in (3+1) oscillation schemes.  Furthermore, the differences
in $\chi^2$ between the best fit point and no oscillations for the two
analyses are given by $\Delta\chi^2_\mathrm{no\,osc} = 29$ (global)
and $\Delta\chi^2_\mathrm{no\,osc} = 47$ (DAR).  This shows that the
information leading to the positive oscillation signal seems to be
more condensed in the DAR data.

Concerning the NEV data, we take into account the detailed information
from the SBL disappearance experiments Bugey~\cite{bugey} and
CDHS~\cite{CDHS}, as described in Ref.~\cite{GS}. The constraints from
KARMEN~\cite{KARMEN} are included by means of the likelihood function
provided by the collaboration~\cite{eitel}. Further, we include the
results on the $\bar\nu_e$ survival probability from the long-baseline
experiments CHOOZ~\cite{CHOOZ} ($P=1.01\pm 0.028\pm 0.027$) 
and Palo Verde~\cite{PaloV} ($P=1.01\pm 0.024 \pm 0.053$)
as explained in Ref.~\cite{Maltoni:2001bc}.

\section{(2+2) oscillations: ruled out by solar and atmospheric data}
\label{sec:2+2} 

The strong preference of oscillations into active neutrinos in solar
and atmospheric oscillations leads to a direct conflict in (2+2)
oscillation schemes \cite{Maltoni:2001bc,peres,Concha2+2}. This is
evident from the comparison of Eqs.~(\ref{eq:etasSol}) and
(\ref{eq:etasAtm}).  We will now show that thanks to the new SNO solar
neutrino data~\cite{sno02} and the improved SK statistic on
atmospheric neutrinos~\cite{skatm} the tension in the data has become
so strong that (2+2) oscillation schemes are, in contrast to
Ref.~\cite{Maltoni:2001bc}, essentially ruled out.

\begin{figure}[t]
  \centering
  \includegraphics[width=0.75\linewidth]{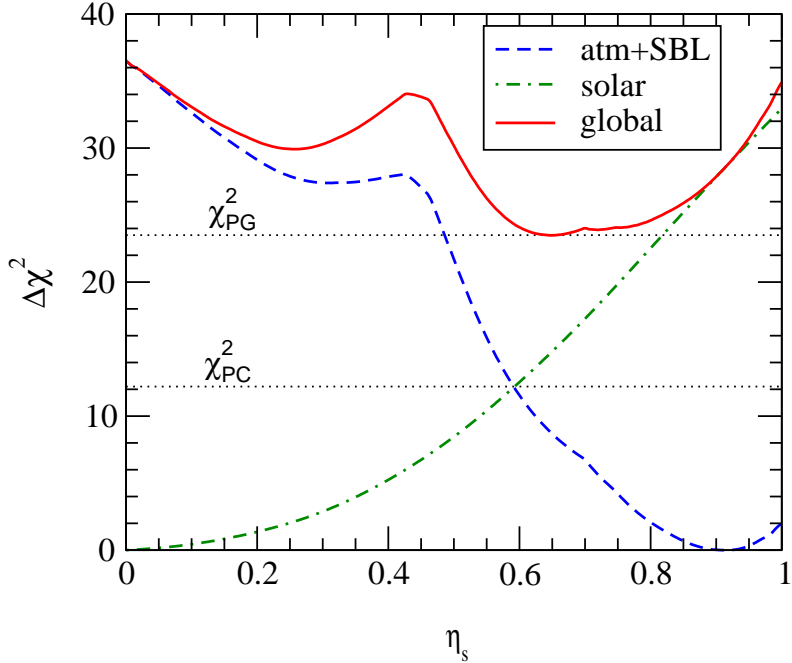}
  \caption{$\Delta\chi^2_\Sol$, $\Delta\chi^2_{\Atm+\Sbl}$
    and $\bar\chi^2_\mathrm{global}$ as a function of $\eta_s$ in
    (2+2) oscillation schemes. Also shown are the values
    $\chi^2_\mathrm{PC}$ and $\chi^2_\mathrm{PG}$ relevant for
    parameter consistency and parameter \gof, respectively.}
  \label{fig:sol+atm}
\end{figure}

In Fig.~\ref{fig:sol+atm} we show the $\chi^2$ as a function of
$\eta_s$ for solar data and for atmospheric combined with SBL
data\footnote{For completeness we use here (atmospheric+NEV+LSND
  global) data.  However, the impact of SBL data on the
  $\chi^2$-dependence on $\eta_s$ is very small, as can be seen by
  comparing the corresponding lines in Figs.~\ref{fig:sol+atm} and
  \ref{fig:atm02.etas}. Also the differences between LSND global and
  LSND DAR are negligible in this case.}.  Furthermore, we show the
$\chi^2$ of the global data defined by
\begin{equation}\label{chi2solatm}
\bar\chi^2(\eta_s) \equiv 
\Delta\chi^2_\Sol(\eta_s) + 
\Delta\chi^2_{\Atm + \Sbl}(\eta_s) \,.
\end{equation}
In this section we investigate quantitatively the degree of
disagreement of the two data sets solar and atmospheric+SBL in (2+2)
oscillation schemes using the statistical methods described in detail
in the appendix.

First, in the method which we are calling {\sl parameter consistency
  test} (PC) we calculate the \CL\ at which still some value of the
common parameter $\eta_s$ is allowed by both data sets. From
Fig.~\ref{fig:sol+atm} we find
\begin{equation}\label{2+2PC}
\chi^2_\mathrm{PC} = 12.2 \,.
\end{equation}
This implies that only if we take both data sets at the 99.95\% \CL\ 
a value of $\eta_s$ exists, which is contained in the allowed regions
of both sets. 

Second, the {\sl parameter goodness of fit} (PG) makes use of the
$\bar\chi^2$ defined in Eq.~(\ref{chi2solatm}). This criterion
evaluates the \gof\ of the {\sl combination} of data sets, without
being diluted by a large number of data points, as it happens for the
usual \gof\ criterion. For the case of two data sets constraining one
single common parameter this method is equivalent to subtracting the
best fit values for the parameter obtained from the two data sets and
calculating the \CL\ at which the difference is consistent with zero.
According to Eq.~(\ref{PG}) we find
\begin{equation}\label{2+2PG}
\chi^2_\mathrm{PG} \equiv \bar\chi^2_\mathrm{min} = 23.5 \,.
\end{equation}
Evaluated for 1 \dof\footnote{Applying Eq.~(\ref{PG}) 
we find for the number of \dof\ 
$3_\Sol + 7_{\Atm + \Sbl} - 9 = 1$ corresponding to the
single common parameter $\eta_s$.}
such a high $\chi^2$-value leads to the marginal 
parameter \gof\ of $1.3 \times 10^{-6}$. 

From Eqs.~(\ref{2+2PC}) and (\ref{2+2PG}) we conclude that (2+2)
oscillation schemes are highly disfavoured, if not ruled out: solar
data and atmospheric+SBL data are in disagreement at the 3.5$\sigma$
level according to the PC test and at the 4.8$\sigma$ level according
to the PG. This is a very robust result, independent of whether LSND
is confirmed or disproved. We note, however, that in the unlikely
event that the KamLAND reactor experiment \cite{kamland} will see no
signal and rule out the LMA solution, the fit of (2+2) schemes may
improve. In that case, although the solar fit quality will
substantially deteriorate since we are left with non-LMA solar
solutions, the agreement with atmospheric data will improve,
especially for the quasi-vacuum case \cite{solat02}.

\section{(3+1) oscillations: strongly disfavoured by SBL data}
\label{sec:3+1}

It is known for a long time that (3+1) mass schemes are disfavoured by
the comparison of SBL disappearance data with the LSND
result~\cite{3+1early,BGGS}. Here we extend the first quantitative
analyses of Refs.~\cite{GS,Maltoni:2001mt} by evaluating whether the
two LSND data sets (global~\cite{LSND2001} and
DAR~\cite{Church:2002tc}) are consistent with all the rest of the
global sample of oscillation data\footnote{Here the NEV data plays a
  key role, but also atmospheric data is important in the lower
  regions of $\dml$ \cite{BGGS,Maltoni:2001mt}. For completeness we
  include also solar data, although the impact on the LSND amplitude
  is very small.}  (SOL+ATM+NEV) by using the statistical methods
explained in detail in the appendix.

\begin{figure}[t]
  \centering 
  \includegraphics[width=0.75\linewidth]{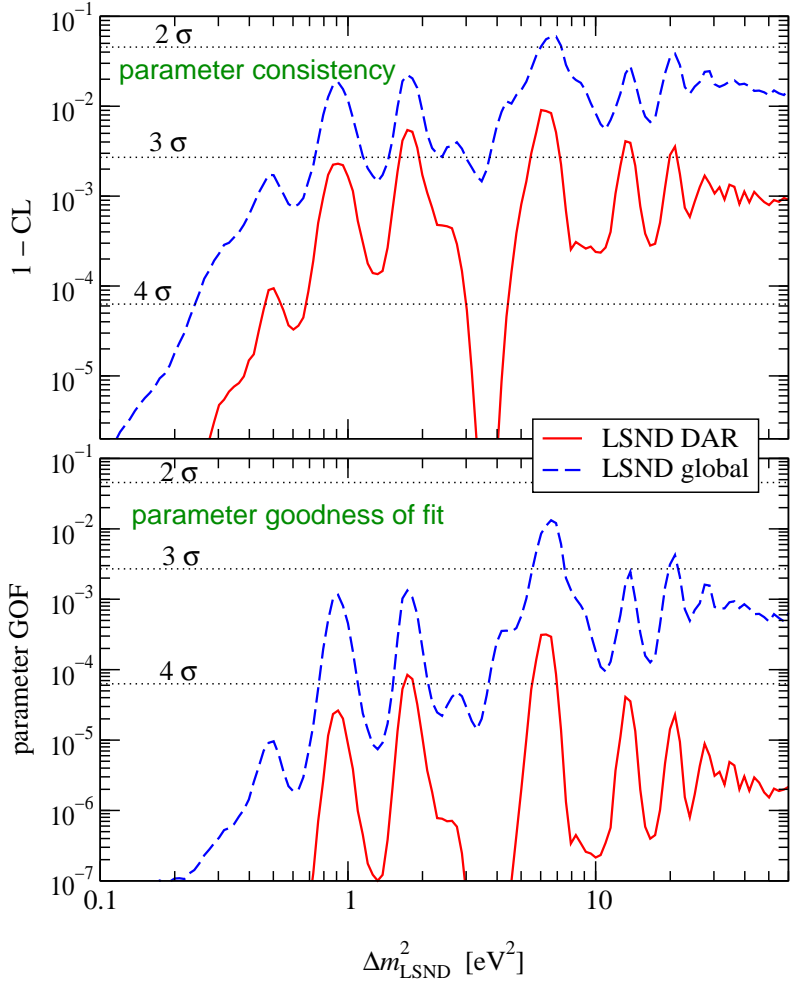}
  \caption{Compatibility of LSND with solar+atmospheric+NEV data in
    (3+1) schemes. In the upper panel we show the \CL\ of the
    parameter consistency whereas in the lower panel we show the
    parameter \gof\ for fixed values of $\dml$. The analysis is
    performed both for the global \cite{LSND2001} and for the DAR
    \cite{Church:2002tc} LSND data samples.}
  \label{fig:3+1}
\end{figure}

Following previous works~\cite{GS,Maltoni:2001mt} and the original
analyses of Bugey~\cite{bugey} and CDHS~\cite{CDHS} we perform the
analysis for {\sl fixed} values of $\dml$. Therefore, the two data
sets we are comparing depend on $1_\mathrm{LSND}$ and
$8_\mathrm{SOL+ATM+NEV}$ parameters and have only the single
parameter $\thl$ in common. For each value of $\dml$ the analysis is
completely analogous to the one of the previous section.  In
Fig.~\ref{fig:3+1} we show the results of the PC 
test and the PG test for both LSND
data sets as a function of $\dml$. The horizontal lines indicate the
\CL\ of 95.45\%, 99.73\% and 99.9937\%, corresponding to 2, 3 and 4
standard deviations, respectively.  
From this figure we observe that the disagreement is roughly one
standard deviation worse with the LSND DAR analysis compared to the
LSND global data analysis. The larger values of $\sin^22\thl$
preferred by the DAR data are in stronger disagreement with the bounds
of the NEV experiments.

The upper panel of Fig.~\ref{fig:3+1} shows that for the global LSND
data a value of $\thl$ consistent with the rest of the oscillation
data exists at best at the 2$\sigma$ level. For LSND DAR parameter
consistency is possible only at a \CL\ greater than 99\% (for most
$\dml$ values, only at more than 3$\sigma$).  The parameter \gof\ is
shown in the lower panel. For the global LSND data it is always worse
than 3$\sigma$, except for $\dml\sim 6$ eV$^2$ it reaches 1\%.  For
LSND DAR it is for most values of $\dml$ below 0.01\%.
The weakest disagreement in all cases occurs for the ``LSND island''
around $\dml\sim 6$ eV$^2$.  Note that this does not correspond to
the best fit value in a global fit (see next section). For the LSND
global data the best fit point is at $\dml=27$ eV$^2$ with a PC at
97\% \CL\ and a PG 0.2\%. For LSND DAR the best fit point is at
$\dml=0.91$ eV$^2$ with a PC at 3$\sigma$ and an PG of $3\times
10^{-5}$.

\section{Comparing (3+1), (2+2) and (3+0) hypotheses}
\label{sec:comparing}

With the methods developed in Ref.~\cite{Maltoni:2001bc} we are able
to perform a global fit to the oscillation data in the four-neutrino
framework. This approach allows to statistically compare the
different hypotheses. Let us first evaluate the \gof\ of (3+1) and
(2+2) spectra with the help of the PG method described in the
appendix. We divide the global oscillation data in the 4 data sets
SOL, ATM, NEV and LSND. According to Eq.~(\ref{PG}) we consider the
$\chi^2$-function 
\begin{eqnarray}
\bar\chi^2 &=& \Delta\chi^2_\Sol(\theta_\Sol,\dms,\eta_s) +
\Delta\chi^2_\Atm(\theta_\Atm,\dma,\eta_s,d_\mu) \nonumber\\
&+& \Delta\chi^2_\Nev(\theta_\Lsnd,\dml,d_\mu,\eta_e) +
\Delta\chi^2_\Lsnd(\theta_\Lsnd,\dml) \,,
\end{eqnarray}
and the relevant number of \dof\ is given by
\begin{equation}
3_\mathrm{SOL} + 4_\mathrm{ATM} + 4_\mathrm{NEV} + 2_\mathrm{LSND}
-9_\mathrm{parameters} = 4 \,.
\end{equation}
This corresponds to the 4 parameters $\eta_s, d_\mu, \thl, \dml$ describing
the coupling of the different data sets.

\begin{table}[t]\centering
    \catcode`?=\active \def?{\hphantom{0}}
    \begin{tabular}{|l|cc|cc|}
    \hline\hline 
    & \multicolumn{2}{c|}{LSND global} &  \multicolumn{2}{c|}{LSND DAR} \\ 
    \hline 
    & (3+1) & (2+2) & (3+1) & (2+2) \\
    \hline
    SOL   & ?0.0 & 14.8 & ?0.0 & 14.8 \\
    ATM   & ?0.4 & ?6.7 & ?0.2 & ?6.7 \\
    NEV   & ?7.0 & ?9.7 & 17.1 & 12.2 \\
    LSND  & ?7.2 & ?1.2 & ?6.8 & ?1.9 \\
    \hline $\chi^2_\mathrm{PG}$
          & 14.6 & 32.4 & 24.1 & 35.6 \\
    \hline
    PG    & $5.6\times 10^{-3}$ & $1.6\times 10^{-6}$ & 
            $7.6\times 10^{-5}$ & $3.5 \times 10^{-7}$  \\
    \hline\hline
    \end{tabular}
    \caption{\label{tab:pg}%
$\chi^2_\mathrm{PG}$ for (3+1) and (2+2) oscillation 
schemes in the global analysis for the two LSND data sets global 
\cite{LSND2001} and DAR \cite{Church:2002tc}. In the last row the 
parameter \gof\ is shown for 4 \dof.}
\end{table}
    
The results of this analysis is shown in Tab.~\ref{tab:pg}. We show
the contributions of the 4 data sets to $\chi^2_\mathrm{PG} \equiv
\bar\chi^2_\mathrm{min}$ for (3+1) and (2+2) oscillation schemes and
for both LSND analyses. As expected we observe that in (3+1) schemes
the main contribution comes from SBL data due to the tension between
LSND and NEV data in these schemes.  For (2+2) oscillation schemes a
large part of $\chi^2_\mathrm{PG}$ comes from solar and atmospheric
data, however, also SBL data contributes significantly.  This comes
mainly from the tension between LSND and KARMEN \cite{Church:2002tc},
which does not depend on the mass scheme and, hence, also contributes
in the case of (2+2).  Therefore, the values of $\chi^2_\mathrm{PG}$
given in Tab.~\ref{tab:pg} for (2+2) schemes are higher than the one
of Eq.~(\ref{2+2PG}), where the tension in SBL data is not included.

In the last row of Tab.~\ref{tab:pg} we show the PG for the various cases.
The best \gof\ is obtained with the global LSND data in the (3+1) scheme. 
However, even in this best case the \gof\ is only 0.56\%. It is fair to
say that all other cases are essentially ruled out by the PG criterion. 
In general the fit is worse with the LSND DAR analysis. Due to the
larger values of $\sin^22\thl$ preferred by this data the disagreement 
between LSND and NEV experiments is significantly stronger than with 
the global data. We observe that the fit gets worse, if one selects
LSND data with higher sensitivity to the oscillation signal. This seems
rather discouraging for the four-neutrino hypothesis as an interpretation 
of LSND.

\begin{table}[t]\centering
    \catcode`?=\active \def?{\hphantom{0}}
    \begin{tabular}{|l|cc|}
    \hline\hline 
    $\Delta \chi^2$ & (2+2) & (3+0) \\
    \hline
    LSND global & 17.8 & 20.0 \\
    LSND DAR    & 11.5 & 28.5 \\
    \hline\hline
    \end{tabular}
    \caption{\label{tab:comparison}%
    $\Delta \chi^2$ of (2+2) and (3+0) oscillation schemes with 
respect to (3+1) spectra  for the two LSND data sets global 
\cite{LSND2001} and DAR \cite{Church:2002tc}.}
\end{table}

Although we have seen that none of the four-neutrino mass schemes can
provide a reasonable good fit to the global oscillation data including
LSND, it might be interesting to consider the {\sl relative} status of
the three hypotheses (3+1), (2+2) and the three-active neutrino
scenario (3+0). This can be done by comparing the $\chi^2$ values of
the best fit point (which is in the (3+1) scheme) with the one
corresponding to (2+2) and (3+0). In Tab.~\ref{tab:comparison} we show
the $\Delta \chi^2$ of (2+2) and (3+0) with respect to (3+1) for LSND
global and DAR data.

First we observe that for LSND global (2+2) is strongly disfavoured
with respect to (3+1) with a $\Delta \chi^2 = 17.8$. The reason for
the big change with respect to the value of $\Delta \chi^2 = 3.7$
found in Ref.~\cite{Maltoni:2001bc} is the improved sensitivity of
solar (SNO NC) and atmospheric (SK 1489-days) data to a sterile
component. With this new data now (3+1) schemes are clearly preferred
over (2+2): for 4 \dof\ a $\Delta \chi^2 = 17.8$ implies that (2+2) is
ruled out at 99.87\% \CL. Further we see from
Tab.~\ref{tab:comparison} that now (2+2) is only slightly better than
(3+0) for LSND global.
Using LSND DAR data (2+2) is somewhat less disfavoured with respect to
(3+1). The reason for this is that for LSND DAR (3+1) gets much worse
compared to LSND global, whereas the fit of (2+2) deteriorates only
slightly (see Tab.~\ref{tab:pg}). Hence, the difference becomes
smaller. Still, (3+1) is clearly preferred: for 4 \dof\ a $\Delta
\chi^2 = 11.5$ implies that (2+2) is ruled out at 97.9\% \CL\ with
respect to (3+1). Of course, for LSND DAR (3+0) is very strongly
disfavoured with a $\Delta \chi^2 = 28.5$ because of the high
significance of the oscillation signal in this data set, which cannot
be accounted for in the (3+0) case.

\section{Conclusions}
\label{sec:conclusions}

Prompted by the recent SNO neutral current result and recent
atmospheric data, we have re-analyzed the global status of all current
neutrino oscillation data in the framework of four-neutrino schemes,
in a way similar to that of Ref.~\cite{Maltoni:2001bc}. Our present
update includes the global solar neutrino data, the most recent sample
of atmospheric neutrino experiments, and a detailed treatment of
experiments reporting no evidence (NEV) for oscillations (KARMEN,
Bugey, CDHS as well as information from CHOOZ and Palo Verde).
Besides the analysis of the global LSND data, we also perform an
analysis based mainly on the decay-at-rest (DAR) data sample, which
provides a higher sensitivity to the oscillation signal in LSND.
We have identified the parameter consistency (PC) test and the parameter
goodness of fit (PG) as useful statistical methods to evaluate the
compatibility of different data sets in a global analysis.

We have found that in (2+2) mass schemes recent solar and atmospheric
data are compatible only at the 3.5$\sigma$ according to the PC test,
and at 4.8$\sigma$ according to the PG method.  In (3+1) mass schemes
the disagreement of LSND with the rest of the oscillation data has
been evaluated for both LSND analyses as a function of $\dml$. For the
global LSND data PC is achieved only at 2$\sigma$ and the PG is below
1\% for all values of $\dml$. Using the LSND DAR sample the
disagreement is always stronger: we find a PC only at 99\% \CL\ and
for most values of $\dml$ the PG is worse than 4$\sigma$.

We have evaluated the \gof\ of a global fit in terms of the PG by
dividing the data in SOL, ATM, LSND and NEV samples.  In the best case
we find a PG of only 0.56\%. This value occurs for the (3+1) mass
scheme and the global LSND data. Using the LSND DAR analysis we get a
PG of $7.6\times 10^{-5}$ for (3+1). For (2+2) oscillation schemes the
situation is worse: we find a very bad PG of $1.6\times 10^{-6}$ for
LSND global and $3.5\times 10^{-7}$ for LSND DAR.  Concerning the
relative status of the hypotheses (3+1), (2+2) and the three-active
neutrino case (3+0) we find that for the LSND global data (2+2) and
also (3+0) are strongly disfavoured with respect to (3+1) with a
$\Delta\chi^2= 17.8$ and $\Delta\chi^2= 20.0$, respectively. For LSND
DAR we find for (2+2) a $\Delta\chi^2 = 11.5$ relative to (3+1), and
the high significance of the oscillation signal condensed in the DAR
sample leads to the huge value of $\Delta\chi^2 = 28.5$ of (3+0)
relative to (3+1).

The exclusion of four-neutrino oscillation schemes of the (2+2)-type
is based on the improved sensitivity of solar and atmospheric neutrino
experiments to oscillations into a sterile neutrino, thanks to recent
experimental data. This is a very robust result, independent of
whether the LSND experiment is confirmed or disproved. The exclusion
of (3+1) schemes depends somehow on the used LSND data
sample\footnote{We want to stress that, although (3+1) models
  themselves are not ruled out, they do not offer an acceptable
  framework for a combined description of current oscillation data, if
  LSND is included. In contrast, any model of the (2+2)-type is not
  viable if the gap separating the two pairs of neutrino states is large
  compared to $\dma$.}. 
Furthermore, it heavily relies on the results of
negative SBL experiments, especially on the Bugey and CDHS
disappearance experiments.  Therefore, if LSND should be confirmed by
MiniBooNE, it will be crucial to improve the experimental data on SBL
$\pnu{e}$ and/or $\pnu{\mu}$ disappearance. If the signal in LSND
should indeed stem from oscillations in a (3+1) mass scheme the SBL
$\pnu{e}$ and $\pnu{\mu}$ disappearance amplitude must be just on the
edge of the sensitivity of current experiments.

In summary, the interpretation of the global neutrino oscillation data
including LSND in terms of four-neutrino mass schemes -- in either
(3+1) or (2+2) realizations -- is strongly disfavoured.  In the best
case we obtain a parameter \gof\ of only 0.56\%. The analysis we have
presented brings the LSND anomaly to a theoretically more puzzling
status indeed. We want to note that also introducing more sterile
neutrinos participating in the oscillations is unlikely to
substantially improve the situation \cite{peres}.  If LSND should be
confirmed by the results of the MiniBooNE experiment the situation
will become even more puzzling. Examples of more far-fetched
alternative explanations are the possibility of lepton number
violating muon decay~\cite{Babu:2002ic} or large CPT violation in the
neutrino sector~\cite{CPT}. Such scenarios will be crucially tested at
the upcoming experiments MiniBooNE~\cite{MiniBooNE} and
KamLAND~\cite{kamland}.

\begin{appendix}

\section{Statistical compatibility of different data sets}
\label{appendix}

In the global analysis of oscillation data in the framework of
four-neutrino schemes we are faced with the problem that for a given mass
scheme some data are in conflict with each other. In this appendix we
present two statistical methods to evaluate quantitatively whether
such data sets are consistent with each other.

Suppose a theoretical model characterized by $P$ parameters
$x=(x_1,\ldots,x_P)$ and $K$ statistically independent experiments
(or data sets) described by the theory. Each of these experiments may
consist of many data points; experiment $k$ observes $N_k$ data
points. Furthermore, experiment $k$ may in general depend only on $P_k
\le P$ parameters $x^k$, which is a sub-set of all parameters $x$ of
the full theory, leading to the $\chi^2$-function
\begin{equation}
\chi^2_k(x^k) = (\chi^2_k)_\mathrm{min} + \Delta\chi^2_k(x^k)\,.
\end{equation}
The three terms are distributed with $N_k,\: N_k-P_k, \: P_k$ degrees
of freedom (\dof). The global $\chi^2$ is given by
\begin{equation}\label{glob}
\chi^2_\mathrm{tot}(x) = \sum_{k=1}^K \chi^2_k(x^k) =
\chi^2_\mathrm{min} + \Delta \chi^2(x)\:.
\end{equation}
Usually the goodness-of-fit (\gof) is evaluated by considering the
value of $\chi^2_\mathrm{min}$, which is distributed with $N-P$ \dof, 
where $N=\sum_k N_k$.

Let us assume, that all the experiments considered {\sl separately}
lead to a good fit: $(\chi^2_k)_\mathrm{min} \approx N_k-P_k$.
However, there may be some disagreement between different experiments.
If the number of data points is very large the disagreement may be
completely washed out and the usual \gof\ would be very good, i.e.\ 
$\chi^2_\mathrm{min} \approx N-P$ despite the
disagreement.\footnote{For the global four-neutrino analysis this effect
  was discussed explicitly in Ref.~\cite{Maltoni:2001bc}.}  Here we
propose two methods which are more useful in global analyses with many
data points, like in the four-neutrino case.

\subsection{Parameter consistency test (PC)}

Let us consider the case of two data sets ($K=2$), which have $\bar
P\le P$ parameters $\bar x$ in common. Then at a given \CL\ the
allowed regions for the parameters $\bar x$ according to each data set
are determined by two $\chi^2$ functions $\Delta \chi^2_1(\bar x)$ and
$\Delta \chi^2_2(\bar x)$, where we minimize with respect to the other
parameters.  Now we calculate $\chi^2_\mathrm{PC}$, defined as the
minimal $\Delta \chi^2$ value where there is still some overlap
between the allowed parameter regions of the two data sets.  This
$\chi^2_\mathrm{PC}$ has to be evaluated for $\bar P$ \dof\ to obtain
the \CL\ at which there exists still some values of the parameters
such that both data sets are consistent.

We call this criterion {\sl parameter consistency test}. It can be
applied if the global data can be divided into {\sl two} potentially
conflicting sets.  Similar methods have been used in
Refs.~\cite{GS,Maltoni:2001mt} to evaluate the consistency of LSND and
NEV experiments in (3+1) schemes and in Ref.~\cite{Church:2002tc} to
check the compatibility of LSND and KARMEN.

\subsection{Parameter goodness of fit (PG)}

 Let us write the global $\chi^2$ Eq.~(\ref{glob}) in the following way:
\begin{equation}\label{tot}
\chi^2_\mathrm{tot}(x) = \sum_{k=1}^K (\chi^2_k)_\mathrm{min} +
\sum_{k=1}^K \Delta\chi^2_k(x^k) \,.
\end{equation}
To evaluate the \gof\ we now use only the second term:
\begin{equation}\label{new}
\bar\chi^2(x) \equiv
\sum_{k=1}^K \Delta\chi^2_k(x^k)
 = \bar\chi^2_\mathrm{min} + \Delta \bar\chi^2(x) \,.
\end{equation}
The first term in Eq.~(\ref{tot}) describes the \gof\ of all the
experiments taken alone, whereas $\bar\chi^2_\mathrm{min}$ in
Eq.~(\ref{new}) corresponds to the contribution to the total minimum
$\chi^2_\mathrm{min}$ which stems from the {\sl combination} of the
data sets. The three terms in Eq.~(\ref{new}) are distributed with
$\sum_k P_k,\: \sum_k P_k -P ,\: P$ \dof, respectively. Hence, to
check the \gof\ we just have to evaluate
\begin{equation}\label{PG}
\chi^2_\mathrm{PG} \equiv \bar\chi^2_\mathrm{min}
\quad\mbox{for}\quad 
\sum_{k=1}^K P_k -P\quad\mbox{\dof.} 
\end{equation}
We call this 
criterion {\sl parameter goodness of fit} and in general it is more
stringent than the parameter consistency test (for an illustration see
Fig.~\ref{fig:sol+atm}).

If $M_p \le K$ experiments depend on the parameter $x_p$ we have the
obvious relation $\sum_{k=1}^K P_k = \sum_{p=1}^P M_p$, and the
relevant number of \dof\ given in Eq.~(\ref{PG}) can be written as
$\sum_{p=1}^P M_p -P = \sum_{p=1}^P (M_p -1)$.  From this relation we
see that parameters relevant for only one experiment ($M_p=1$) do not
contribute because they are minimized and contribute only to the
individual $(\chi^2_k)_\mathrm{min}$ which are not considered in
$\chi^2_\mathrm{PG}$.  What is tested is only the parameter
dependence of the data sets relevant for the {\sl combination}.

{\sl A simple example:} Let us consider the case of only one
parameter $y$, which is determined by 2 experiments. Hence, we have
$K=2$ and $P=P_1=P_2=1$ and $\chi^2_\mathrm{PG} = \mbox{min} \left[
  \Delta\chi^2_1(y) + \Delta\chi^2_2(y) \right]$, which has to be
evaluated for 1 \dof.  The following statement is easy to prove: Let
$y_1$ and $y_2$ be the best fit values of the two experiments
($\Delta\chi^2_k(y_k)=0$). Then the \CL\ at which $y_1 - y_2$ is
consistent with zero is identical to the parameter goodness of fit.

\end{appendix}

\section*{Acknowledgments}
We are grateful to W.C.~Louis and G.~Mills for providing us with the
table of the global LSND likelihood function. Further, we thank
K.~Eitel for providing us with the likelihood functions of the LSND
DAR and the KARMEN analyses and for very useful discussions. This work
was supported by Spanish grant PB98-0693, by the European Commission
RTN network HPRN-CT-2000-00148 and by the European Science Foundation
network grant N.~86. T.S.~has been supported by the DOC fellowship of
the Austrian Academy of Science and, in the early stages of this work,
by a fellowship of the European Commission Research Training Site
contract HPMT-2000-00124 of the host group. M.M.~was supported by
contract HPMF-CT-2000-01008 and M.A.T.\ was supported by the M.E.C.D.\
fellowship AP2000-1953.

\end{document}